\documentclass[pra,aps,twocolumn,showpacs,superscriptaddress,amsmath,amssymb]{revtex4-1}
\usepackage{color}

\usepackage{amsmath,graphicx}

\begin{document}
\def\tr{\rm{Tr}}
\def\la{{\langle}}
\def\ra{{\rangle}}
\def\a{{\alpha}}
\def\e{\epsilon}
\def\q{\quad}
\def\w{\tilde{W}}
\def\t{\tilde{t}}
\def\a{\hat{A}}
\def\h{\hat{H}}
\def\E{\mathcal{E}}
\def\p{\hat{P}}
\def\u{\hat{U}}

\title{Bell's Theorem and Entangled Solitons}
%
%
\author {Yu.P. Rybakov}
\affiliation{Theoretical Physics Department, Peoples' Friendship University of Russia, Moscow, Russian Federation}
\author {T.F. Kamalov}
\affiliation{Theoretical Physics Department, Moscow Institute of Physics and Technology, Moscow, Russian Federation}
\affiliation{Department of Physics, Moscow State University of Mechanical Engineering, Moscow, Russian Federation}
\affiliation{Theoretical Physics Department, Skolkovo Institute of Science and Technology, Moscow, Russian Federation}

\date{\today}
\begin{abstract}
Entangled solitons construction being introduced in the nonlinear spinor
field model, the Einstein---Podolsky---Rosen (EPR) spin correlation is
calculated and shown to coincide with the quantum mechanical one for the $%
1/2 $--spin particles.

\end{abstract}

%
%
\maketitle

\section[Introduction]{Introduction}

\label{c:1} According to L.~de~Broglie~[1] and A.~Einstein~[2], particles
are considered as spatial regions with very high intensity of some
fundamental field. Nowadays such field configurations are known as solitons.
In this article we find new arguments in favour of the thought that the
soliton concept can give a consistent description of extended quantum
particles.

First of all we recall that the famous Bell's theorem~[3] states that
hypothetical hidden variables in quantum mechanics cannot be considered as
local (point-like) ones. However, we intend to show that solitons prove to
be considered as non-local (extended) hidden variables.

\section{Entangled solitons and EPR correlations}

\label{c:2}

In the sequel we shall consider the special case of two--particles
configurations corresponding to the singlet state of two $1/2$--spin
particles. In quantum mechanics these states are described by the spin wave
function of the form
\begin{equation}  \label{eq:1}
\psi _{12} = \frac{1}{\sqrt{2}}\left(|1\,\uparrow \rangle \otimes
|2\,\downarrow \rangle - |1\,\downarrow \rangle \otimes |2\,\uparrow \rangle
\right)
\end{equation}
and are known as \textbf{entangled states}. The arrows in (\ref{eq:1})
signify the projections of spin $\pm 1/2$ along some fixed direction. In the
case of the electrons in the famous Stern---Gerlach experiment this
direction is determined by that of an external magnetic field. If one
chooses two different Stern---Gerlach devices, with the directions $\mathbf{a%
}$ and $\mathbf{b}$ of the magnetic fields, denoted by the unit vectors $%
\mathbf{a}$ and $\mathbf{b}$ respectively, one can measure the correlation
of spins of the two electrons by projecting the spin of the first electron
on $\mathbf{a}$ and the second one on $\mathbf{b}$. Quantum mechanics gives
for the spin correlation function the well-known expression
\begin{equation}  \label{eq:2}
P(\mathbf{a},\,\mathbf{b}) = \psi _{12}^{+}(\mathbf{\sigma a})\otimes (%
\mathbf{\sigma b})\psi_{12},
\end{equation}
where $\mathbf{\sigma }$ stands for the vector of Pauli matrices $\sigma _i$%
, $i = 1, 2, 3$. Putting (\ref{eq:1}) into (\ref{eq:2}), one easily gets
\begin{equation}  \label{eq:3}
P(\mathbf{a},\,\mathbf{b}) = -(\mathbf{ab}).
\end{equation}

The formula (\ref{eq:3}) characterizes the spin correlation in the
Einstein---Podolsky---Rosen entangled singlet states and is known as the
EPR--correlation. As was shown by J.~Bell~[3], the correlation (\ref{eq:3})
can be used as an efficient criterium for distinguishing the models with the
local (point-like) hidden variables from those with the non-local ones.
Namely, for the local-hidden-variables theories the EPR--correlation (\ref%
{eq:3}) is broken.

It would be interesting to check the solitonian model, shortly described in
the beforehand points, by applying to it the EPR--correlation criterium. To
this end let us first describe the $1/2$--spin particles as solitons in the
nonlinear spinor model of Heisenberg---Ivanenko type considered in the
works~[4, 5]. The soliton in question is described by the relativistic
4--spinor field $\varphi $ of stationary type
\begin{equation}
\varphi =\left[
\begin{array}{c}
u \\
v%
\end{array}%
\right] \,e^{-\imath \omega t},  \label{eq:4}
\end{equation}%
satisfying the equation
\begin{equation}
\left( \imath \gamma ^{k}\partial _{k}-\ell _{0}^{-1}+\lambda (\bar{\varphi}%
\varphi )\right) \varphi =0,  \label{eq:5}
\end{equation}%
where $u$ and $v$ denote 2--spinors, $k$ runs Minkowsky space indices 0, 1,
2, 3; $\ell _{0}$ stands for some characteristic length (the size of the
particle---soliton), $\lambda $ is self-coupling constant, $\bar{\varphi}%
\equiv \varphi ^{+}\gamma ^{0}$, $\gamma ^{k}$ are the Dirac matrices. The
stationary solution to the equation (\ref{eq:5}) can be obtained by
separating variables in spherical coordinates $r$, $\vartheta $, $\alpha $
via the substitution
\begin{equation}
u=\frac{1}{\sqrt{4\pi }}f(r)\left[
\begin{array}{c}
1 \\
0%
\end{array}%
\right] ,\quad v=\frac{i}{\sqrt{4\pi }}g(r)\sigma _{r}\left[
\begin{array}{c}
1 \\
0%
\end{array}%
\right] ,  \label{eq:6}
\end{equation}%
where $\sigma _{r}=(\mathbf{\sigma r})/r$. Inserting (\ref{eq:6}) into (\ref%
{eq:5}), one finds
\begin{gather*}
\frac{\omega }{c}u+\imath (\mathbf{\sigma \bigtriangledown })v-\ell
_{0}^{-1}u+\frac{\lambda }{4\pi }\left( f^{2}-g^{2}\right) u=0, \\
\frac{\omega }{c}v+\imath (\mathbf{\sigma \bigtriangledown })u-\ell
_{0}^{-1}v+\frac{\lambda }{4\pi }\left( f^{2}-g^{2}\right) v=0.
\end{gather*}

In view of (\ref{eq:6}) one gets
\begin{gather*}
\i (\mathbf{\sigma \bigtriangledown })v = -\frac{1}{\sqrt{4\pi }}%
\left(g^{\prime}+ \frac{2}{r}g\right)\left[%
\begin{array}{c}
1 \\
0%
\end{array}%
\right], \\
\i (\mathbf{\sigma \bigtriangledown })u = -\frac{\i }{\sqrt{4\pi }}%
f^{\prime}\sigma _r \left[%
\begin{array}{c}
1 \\
0%
\end{array}%
\right].
\end{gather*}

Finally, one derives the following ordinary differential equations for the
radial functions $f(r)$ and $g(r)$:
\begin{gather*}
\left(g^{\prime}+ \frac{2}{r}g\right) = \left(\frac{\omega }{c} - \ell
_0^{-1}\right)f + \frac{\lambda }{4\pi }\left(f^2 - g^2\right)f, \\
-f^{\prime}= \left(\frac{\omega }{c} + \ell _0^{-1}\right)g + \frac{\lambda
}{4\pi }\left(f^2 - g^2\right)g.
\end{gather*}
As was shown in the papers~[4, 5], these equations admit regular solutions,
if the frequency parameter $\omega $ belongs to the interval
\begin{equation}  \label{eq:7}
0 < \omega < c/{\ell _0}.
\end{equation}

The behavior of the functions $f(r)$ and $g(r)$ at $r\rightarrow 0$ is as
follows:
\begin{equation*}
g(r)=C_{1}r,\quad f=C_{2},\quad f^{\prime }\rightarrow 0,
\end{equation*}%
where $C_{1}$, $C_{2}$ denote some integration constants. The behavior of
solutions far from the center of the soliton, i.e. at $r\rightarrow \infty $%
, is given by the relations:
\begin{equation*}
f=\frac{A}{r}e^{-\nu r},\quad g=-\frac{f^{\prime }}{B},
\end{equation*}%
where
\begin{equation*}
\nu =\left( \ell _{0}^{-2}-{\omega ^{2}}/{c^{2}}\right) ^{1/2},\quad B=\ell
_{0}^{-1}+{\omega }/c.
\end{equation*}

If one chooses the free parameters $\ell _{0}$ and $\lambda $ of the model
to satisfy the normalization condition
\begin{equation}
\int d^{3}x\,\varphi ^{+}\varphi =\int\limits_{0}^{\infty }drr^{2}\left(
f^{2}+g^{2}\right) =\hbar ,  \label{eq:8}
\end{equation}%
then the spin of the soliton reads
\begin{equation}
\mathbf{S}=\int d^{3}x\,\varphi ^{+}\mathbf{J}\varphi =\frac{\hbar }{2}%
\mathbf{e}_{z},  \label{eq:9}
\end{equation}%
where $\mathbf{e}_{z}$ denotes the unit vector along the $Z$--direction, $%
\mathbf{J}$ stands for the angular momentum operator
\begin{equation}
\mathbf{J}=-i[\mathbf{r}\mathbf{\bigtriangledown }]+\frac{1}{2}\,\mathbf{%
\sigma }\otimes \sigma _{0},  \label{eq:10}
\end{equation}%
and $\sigma _{0}$ is the unit $2\times 2$--matrix.

Now it is worth-while to show the positiveness of the energy $E$ of the $1/2$%
--spin soliton. The energy $E$ is given by the expression
\begin{equation}
E=c\,\int d^{3}x\,\left[ -i\varphi ^{+}(\mathbf{\alpha \bigtriangledown }%
)\varphi +\ell _{0}^{-1}\bar{\varphi}\varphi -\frac{\lambda }{2}(\bar{\varphi%
}\varphi )^{2}\right] ,  \label{eq:11}
\end{equation}%
where $\mathbf{\alpha }=\mathbf{\sigma }\otimes \sigma _{1}$. The
positiveness of the functional (\ref{eq:11}) emerges from the virial
identities characteristic for the model in question. In fact, the equation
for the stationary solution (\ref{eq:4}) can be derived from the variational
principle based on the Lagrangian of the system
\begin{equation}
L=-E+\int d^{3}x\,\omega \varphi ^{+}\varphi .  \label{eq:12}
\end{equation}%
Performing the two-parameters scale transformation of the form $\varphi
(x)\rightarrow \alpha \varphi (\beta x)$, one can derive from (\ref{eq:12})
and the variational principle $\delta L=0$ the following two virial
identities, which are valid for any regular stationary solution to the field
equation (\ref{eq:5}):
\begin{gather}
\int d^{3}x\,[i\frac{2}{3}\varphi ^{+}(\mathbf{\alpha \bigtriangledown }%
)\varphi +\frac{\omega }{c}\varphi ^{+}\varphi -\ell _{0}^{-1}\bar{\varphi}%
\varphi +\frac{\lambda }{2}(\bar{\varphi}\varphi )^{2}]=0,  \label{eq:13} \\
\int d^{3}x\,[i\varphi ^{+}(\mathbf{\alpha \bigtriangledown })\varphi +\frac{%
\omega }{c}\varphi ^{+}\varphi -\ell _{0}^{-1}\bar{\varphi}\varphi +\lambda (%
\bar{\varphi}\varphi )^{2}]=0.  \label{eq:14}
\end{gather}

Using (\ref{eq:13}) and (\ref{eq:14}), one can express some sign-changing
integrals through those of definite sign:
\begin{align}
\int d^{3}x[-i\frac{1}{3}\varphi ^{+}(\mathbf{\alpha \bigtriangledown }%
)\varphi ]& =\frac{\lambda }{2}\int d^{3}x\,(\bar{\varphi}\varphi )^{2},
\label{eq:15} \\
\int d^{3}x\,[\ell _{0}^{-1}\bar{\varphi}\varphi +\frac{\lambda }{2}(\bar{%
\varphi}\varphi )^{2}]& =\frac{\omega }{c}\int d^{3}x\,\varphi ^{+}\varphi .
\label{eq:16}
\end{align}%
Using the identities (\ref{eq:15}) and (\ref{eq:16}), one can represent the
energy (\ref{eq:11}) of the soliton as follows:
\begin{equation}
E=c\,\int d^{3}x\,[\ell _{0}^{-1}\bar{\varphi}\varphi +\lambda (\bar{\varphi}%
\varphi )^{2}]=\omega \,\int d^{3}x\,\varphi ^{+}\varphi =\hbar \omega ,
\label{eq:17}
\end{equation}%
where the normalization condition (\ref{eq:8}) was taken into account. Thus,
one concludes, in the connection with (\ref{eq:7}) and (\ref{eq:17}), that
the energy of the stationary spinor soliton (\ref{eq:4}) in the nonlinear
model (\ref{eq:5}) turns out to be positive. Moreover, one can see that (\ref%
{eq:17}) is equivalent to the Planck---de Broglie wave---particle dualism
relation. Now let us construct the two--particles singlet configuration on
the base of the soliton solution (\ref{eq:4}). First of all, in analogy with
(\ref{eq:1}), one constructs the entangled solitons configuration endowed
with the zero spin:
\begin{equation}
\varphi _{12}=\frac{1}{\sqrt{2}}\left[ \varphi _{1}^{\uparrow }\otimes
\varphi _{2}^{\downarrow }-\varphi _{1}^{\downarrow }\otimes \varphi
_{2}^{\uparrow }\right] ,  \label{eq:18}
\end{equation}%
where $\varphi _{1}^{\uparrow }$ corresponds to (\ref{eq:6}) with $\mathbf{r}%
={\mathbf{r}}_{1}$, and $\varphi _{2}^{\downarrow }$ emerges from the above
solution by the substitution
\begin{equation*}
{\mathbf{r}}_{1}\rightarrow {\mathbf{r}}_{2},\quad \left[
\begin{array}{c}
1 \\
0%
\end{array}%
\right] \rightarrow \left[
\begin{array}{c}
0 \\
1%
\end{array}%
\right] \,
\end{equation*}%
that corresponds to the opposite projection of spin on the $Z$--axis. In
virtue of the orthogonality relation for the states with the opposite spin
projections, one easily derives the following normalization condition for
the entangled solitons configuration (\ref{eq:18}):
\begin{equation}
\int d^{3}x_{1}\,\int \,d^{3}x_{2}\,\varphi _{12}^{+}\varphi _{12}={\hbar }%
^{2}.  \label{eq:19}
\end{equation}

Now it is not difficult to find the expression for the stochastic wave
function ~[6--8] for the singlet two--solitons state:
\begin{equation}  \label{eq:20}
\Psi _N \left(t, {\mathbf{r}}_1, {\mathbf{r}}_2 \right) = {\left({\hbar }%
^{2}N\right)}^{-1/2}\sum\limits_{j=1}^{N}\,\varphi_{12}^{(j)},
\end{equation}
where $\varphi_{12}^{(j)}$ corresponds to the entangled soliton
configuration in the $j$--th trial, with the number of trials $N \gg 1$.

Our final step is the calculation of the spin correlation (\ref{eq:2}) for
the singlet two--soliton state. In the light of the fact that the operator $%
\mathbf{\sigma }$ in (\ref{eq:2}) corresponds to the twice angular momentum
operator (\ref{eq:10}), one should calculate the following expression:
\begin{equation}
P^{\prime }(\mathbf{a},\mathbf{b})=\mathbb{M}\,\int d^{3}x_{1}\,\int
\,d^{3}x_{2}\,\Psi _{N}^{+}2\left( {\mathbf{J}}_{1}\mathbf{a}\right) \otimes
2\left( {\mathbf{J}}_{2}\mathbf{b}\right) \Psi _{N},  \label{eq:21}
\end{equation}%
where $\mathbb{M}$ stands for the averaging over the random phases of the
solitons. Inserting (\ref{eq:20}) and (\ref{eq:10}) into (\ref{eq:21}),
using the independence of trials $j\not=j^{\prime }$ and taking into account
the relations:
\begin{align*}
J_{+}\varphi ^{\uparrow }& =0, & J_{3}\varphi ^{\uparrow }& =\frac{1}{2}%
\varphi ^{\uparrow }, & J_{-}\varphi ^{\uparrow }& =\varphi ^{\downarrow },
\\
J_{-}\varphi ^{\downarrow }& =0, & J_{3}\varphi ^{\downarrow }& =-\frac{1}{2}%
\varphi ^{\downarrow }, & J_{+}\varphi ^{\downarrow }& =\varphi ^{\uparrow },
\end{align*}%
where $J_{\pm }=J_{1}\pm :iJ_{2}$, one easily finds that
\begin{equation}
P^{\prime }(\mathbf{a},\mathbf{b})=-{\hbar }^{-2}\left( \mathbf{a}\mathbf{b}%
\right) \left( \int\limits_{0}^{\infty }d\,rr^{2}\left( f^{2}+g^{2}\right)
\right) ^{2}=-\left( \mathbf{a}\mathbf{b}\right) .  \label{eq:22}
\end{equation}%
Comparing the correlations (\ref{eq:22}) and (\ref{eq:3}), one remarks their
coincidence, that is the solitonian model satisfies the EPR--correlation
criterium.

\section{Conclusion}

\label{c:3} The coincidence of the quantum spin correlation with that in the
solitonian scheme supports the hope that the latter one has many attractive
features relevant to consistent theory of extended elementary
particles~[9--11].

\end{document}